\documentclass{ws-rv975x65}
\usepackage{subfigure}     
\usepackage{ws-rv-thm}     
\usepackage[square]{ws-rv-van}
\makeindex

\def\dd{{\rm d}}
\def\f{\frac}
\def\B{{}_{\rm B}}

\newcommand{\be}{\begin{equation}}
\newcommand{\ee}{\end{equation}}
\newcommand\ham{{\cal H}}
\newcommand\mub{\bar{\mu}}
\newcommand\sQ{{\sigma_{Q}^2}}

\begin{document}

\chapter[Loop quantum gravity and observations]{Loop quantum gravity and observations}\label{ra_ch1}

\author[F. Author and S. Author]{Aur\'elien Barrau}

\address{Laboratoire de Physique Subatomique et de Cosmologie,\\
Universit\'e Grenoble-Alpes, CNRS-IN2P3,\\
53 avenue des Martyrs, 38026 Grenoble cedex, France\\
Aurelien.Barrau@cern.ch}

\author[F. Author and S. Author]{Julien Grain}

\address{Institut d'Astrophysique Spatiale, Universit\'e Paris-Sud 11, CNRS \\ B\^atiments 120-121, 91405
Orsay Cedex, France\\
julien.grain@ias.u-psud.fr}

\begin{abstract}
Quantum gravity has long been thought to be completely decoupled from experiments or observations. Although it is true that smoking guns are still missing, there are now serious hopes that quantum gravity phenomena might be tested. We review here some possible ways to observe loop quantum gravity effects in cosmology and astroparticle physics. \\

{\it Invited review article for the World Scientific book "Loop Quantum Gravity" edited by A. Ashtekar and J. Pullin.}
\end{abstract}
\body

\section{Introduction}\label{ra_sec1}
Building a quantum theory of space-time might be the most outstanding problem of contemporary fundamental theoretical physics. Probably this is not mainly because unification is necessary and unavoidable. Unification is unquestionably a useful guide that has indeed helped a lot in the past but that might very well not be the final word on what physics should look like. After all, it could be that different sub fields of physics are described by different theories. The key issue has more to do with consistency. In some circumstances quantum mechanics and string gravity are simultaneously acting. In addition, the quantum world has interactions with the gravitational field, which automatically requires gravity to be understood in a quantum language, as can be demonstrated by appropriate thought experiments. Furthermore, the nonlinearity of gravity frustrates all attempts to ignore quantum gravity: as soon as a strong gravitational field is involved, the coupling
to gravitons should also be strong. The very existence of singularities in general relativity (GR) also requires a quantum extension. Finally, even if the signal measured by the BICEP2 experiment \cite{bicep2} was not from cosmological origin, there is a reasonable hope that primordial gravitational waves will be soon seen through B-modes in the cosmological microwave background: this would be the first observation of a quantum gravity phenomenon, at a linear level though, in the history of science.

Several non-perturbative and background-independent approaches have been developed in the last decades. Among them, loop quantum gravity (LQG) might be the most advanced one (see \cite{lqg_review}  for introductions). One of the main achievements of LQG is to be able to lead to experimental predictions. At this stage, none of them has yet been tested and some of them are still controversial. There are even tensions between different approximation schemes within LGQ. Still, it is a remarkable achievement that a quantum theory of gravity is now able to produce a set of predictions that might be tested in a quite near future.

In this brief review, we will first focus on cosmology, considering different probes, both direct and indirect. We will then consider possible consequences of a possible Lorentz invariance violation. Evaporating black holes will be studied and, finally, we will mention new ideas about Planck stars.

\section{Cosmology: indirect probes}\label{ra_sec2}
When assumed to be isotropic and homogeneous, the Universe is symmetric enough to be a quite easy system to quantize. As explained in the chapter of this book written by Agullo and Singh, and as reviewed in \cite{lqc_review}, LQG ideas have been successfully applied to this specific situation: this is what loop quantum cosmology (LQC) describes. Although a rigorous derivation of LQC from LQG is still missing, it is now fairly believed to capture effectively most quantum effects from the mother theory. Recent progresses were, {\it e.g.}, reported in \cite{alesci}. The most important result is probably the singularity resolution: the Big Bang is replaced by a Big Bounce and the LQC dynamics is different from the Wheeler--de Witt one.

It is difficult but possible to make predictions for perturbations in LQC. Two main paths are followed at this stage. On the one hand  a ``dressed metric approach" \cite{aan} was developed. It tries to deal deeply with quantum fields on a quantum background geometry. On the other hand, an "effective approach"  \cite{effective} was investigated. It tries to avoid fixing or assuming any background structure but instead derives it from effective equations. Both deserve to be seriously considered.

In this section we therefore first focus on more "reliable" predictions related to the background evolution. Holonomy corrections appear in the theory because there is no operator associated with the Ashtekar connexion but only with its holonomy. Although the way those corrections are implemented, leading to the bounce, can of course be questioned, the main picture is now consistent and well established. 

\subsection{Isotropic case}
\subsubsection{Initial conditions at the bounce}
A first approach, developed in \cite{abhay}, assumes that the bounce is the appropriate time to set initial conditions. This is reasonable as the bounce is the only specific point in time. The Universe is also, as usually done in inflation, assumed to be filled with a massive scalar field.

The idea is to solve, thanks to the bounce, the ambiguity that usually appears in the construction of a measure on the space of initial data. The space of solutions  is  isomorphic to a gauge fixed surface, {\it i.e.,} a 2-surface
$\hat\Gamma$ which is intersected by each dynamical
trajectory only once. Since $b$, the conjugate momentum to the volume of the fixed fiducial cell used in the quantization, is monotonic in each solution, the strategy is to choose for $\hat\Gamma$ an appropriate
2-surface $b= b_o$. Symplectic geometry considerations unambiguously equip $\hat\Gamma$ with an induced
Liouville measure $\dd\hat\mu_{\mathcal{L}}$. 
A natural choice is to set $b_o = \pi/2\lambda$ so that $\hat\Gamma$ is naturally coordinatized by
$({\bar\varphi}_{B},v_{B})$, the scalar field and the volume at the bounce.
The induced
measure is given by
$ \dd\hat\mu_{\mathcal{L}} =   \f{\sqrt{3\pi}}{\lambda}\,\, \big[1-
x_{B}^2\big]^{\f{1}{2}}\,\, {\dd} {\bar\varphi}_{\B}\, {\dd} {v}_{\B}$, where $x_B^2$ is the value of $x^2$ 
at the bounce (with $x^2=m^2 {\bar\varphi}^2/(2 \rho_c))$, that is the fraction of  total energy
density in form of potential energy at the bounce. After factoring out the gauge orbits the
 fractional volumes of physically relevant sub-regions of $\hat\Gamma$ can be calculated.
The main results of the study performed in \cite{abhay}, depending on 3 different possible regimes, are: \\
$\bullet$ for $x_B^2 < 10^{-4}$, the number of
e-folds during  slow roll is given approximately by
$ N \approx 2\pi\, \big(1-\f{{\bar\varphi}_o^2}{{\bar\varphi}_{\rm
max}^2}\big)\, {\bar\varphi}_o^2\, \ln {\bar\varphi}_o,$
where ${\bar\varphi}_o$ is the value of the scalar field at the onset
of inflation and ${\bar\varphi}_{\rm max} = 1.5 \times 10^{6}$. For ${\bar\varphi}_{B} = 0.99$, one has ${\bar\varphi}_o = 3.24$
and $N = 68$. Thus, there is a slow roll inflation with over $68$ e-foldings for all
${\bar\varphi}_{B}>1$, {\it i.e.}, if $x_B^2 > 4.4\times 10^{-13}$. \\
$\bullet$ for $10^{-4} < x_B^2 < 0.5$, the LQC departures from GR are
now  significant. 
%
%
The Hubble parameter is 
essentially frozen at a very high value. Throughout this range of $x_B^2$ there are more than 68 e-foldings. \\
$\bullet$ for $0.5 <x_B^2 < 1$, the  LQC effects strongly dominate. Again, because $\dot{\bar\varphi}
>0$, the inflaton climbs up the potential but  the turn around
($\dot{\bar\varphi} =0$) 
occurs during super-inflation.  The Hubble parameter freezes at the onset of inflation and the slow roll conditions are easily met as $\dot{H}/H^2$
is less than $1 \times 10^{-11}$ when $\ddot{\bar\varphi} =0$. There are many more
than 68 e-foldings already in the super-inflation phase. The
friction term is large and the inflation enters a long (more than 68 e-folds) slow roll
inflationary phase. 

Basically all LQC dynamical trajectories are funneled to conditions
which virtually guarantee slow-roll inflation with more than 68
e-foldings, without any input from the pre-big bang regime.  This work was developed further, using analytical and numerical
methods, to calculate the {\it a priori} probability of
realizing a slow-roll phase compatible with CMB. It was
found that the probability is greater than 0.999997 in LQC. This can be considered as a good indirect  -- although not definitive -- test of LQC.

\subsubsection{Initial conditions in the remote past}

In \cite{abhay},  the probability distribution is assumed to be flat and defined at the bounce 
(the first attempts in this direction were performed in \cite{first}). It is however possible to make 
a very different assumption: the phase of the oscillations of the field in the remote past can also be considered 
as a very natural random variable \cite{linda1}. The choice of what is a natural 
measure depends heavily on when 
one decides to set initial conditions  \cite{Corichi:2010zp}. It is important to consider seriously the meaning of an 
``initial'' condition in a Universe that has a contracting branch before the bounce.  In this approach one does not the focus on the initial data at the bounce as in \cite{abhay}, but rather derives a probability  distribution for them as a prediction of the model.

The approach consists in calculating the probability distribution for $x_B$, the square root of the fraction 
of potential energy at the bounce, and $N$, the number of e-folds of slow-roll inflation. 
The most natural and consistent assumption is to set the initial probability 
distribution in the pre-bounce oscillatory phase where the Universe is in addition classical and therefore well under control. The evolution in this phase is described by:
$
\rho=\rho_0\left(1-\frac{1}{2}\sqrt{3\kappa\rho_0}\left( t+\frac{1}{2m}\sin(2mt+2\delta)\right)\right)^{-2},\label{prerho}
$ with
$
x=\sqrt{\frac{\rho}{\rho_c}}\sin(mt+\delta)~,~ y=\sqrt{\frac{\rho}{\rho_c}}\cos(mt+\delta).  \label{prex}
$
In fact, due to hidden symmetries, $\delta$ can be shown to be the only parameter. 

\begin{figure}
\begin{center}
\includegraphics[width=100mm]{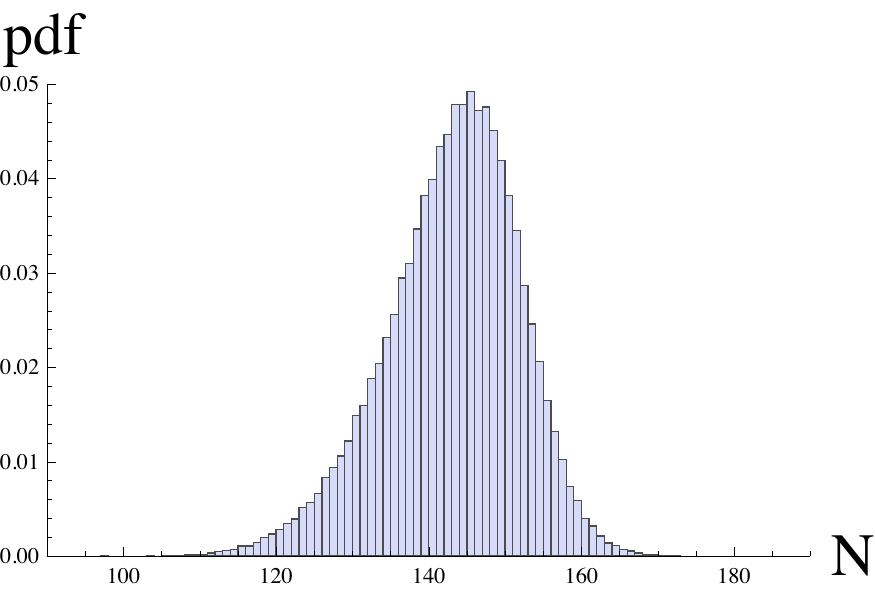}
\caption{Probability distribution function of the number of e-folds of slow-roll inflation (from \cite{linda1}).}
\label{N}
\end{center}
\end{figure}

In addition of being the obviously expected distribution for any oscillatory process of this kind, a flat 
probability for $\delta$ will be preserved over time during the pre-bounce oscillations, making it 
a very natural choice. This is not a trivial point as any other probability distribution 
would be distorted over time, meaning that the final result in the full numerical analysis would depend 
on the choice of $\rho_0$. Starting with a flat probability distribution for $\delta$, the probability for different values of $x_B$ 
can be calculated numerically. In \cite{abhay}, $x_B$ is considered as unknown whereas, in this second approach \cite{linda1}, it is shown to be sharply peaked around $3.55\times 10^{-6}$ (this value scales with $m$ as $m\log\left(\frac{1}{m}\right)$, 
where we assumed that $m\ll1$ in Plank units). The most likely solutions are exactly those that have no slow-roll deflation. 
The probability density for $N$ can also be computed and is given in 
Fig. \ref{N}, showing that the model leads to a slow-roll inflation of about 140 e-folds. This becomes, as shown in \cite{linda1}, a 
{\it prediction} of effective LQC: inflation and its duration are not arbitrary anymore.

\subsection{Anisotropic case}
In bouncing cosmologies, either from the loop approach or any other, the question of anisotropies is very important for a  clear reason: the shear term  varies as $1/a^6$ where $a$ is the "mean" scale factor of the Universe. When the Universe is contracting, the shear term becomes more and more important and eventually drives the dynamics. The reason for which the shear can be neglected in standard cosmology is precisely the reason why it becomes important in bouncing models. 
The question of predicting the duration of inflation in LQC was studied in the Bianchi-I case. 

The metric is given by
$
ds^2 := -N^2 d\tau^2 + a_1^2dx^2 + a_2^2dy^2 + a_3^2dz^2,
$
where $a_i$ denotes the directional scale factors. The classical gravitational hamiltonian is
$
\ham_G = \frac{N}{\kappa\gamma^2}\left(\sqrt{\frac{p_1 p_2}{p_3}}c_1c_2+\sqrt{\frac{p_2 p_3}{p_1}}c_2c_2+\sqrt{\frac{p_3 p_1}{p_2}}c_2c_3\right),
\label{classhamG}
$
with Poisson brackets
$
\{c_i,p_j\}=\kappa\gamma\delta_{ij}.
$
The classical directional scale factors can be written as
$
a_1 = \sqrt{\frac{p_2 p_3}{p_1}}
$
and cyclic expressions. The holonomy correction is implemented to account for specific LQG effects with the usual prescription (the framework was introduced in \cite{0903})
$
c_i\rightarrow \frac{\sin(\mub_i c_i)}{\mub_i}.
$
The $\mub_i$ are given by
$
\mub_1 = \lambda\sqrt{\frac{p_1}{p_2p_3}}
$
and cyclic expressions,
where $\lambda$ is the square root of the minimum area eigenvalue of the LQG area operator ($\lambda=\sqrt{\Delta}$). The quantum corrected gravitational Hamiltonian is:
\be
\ham_G=-\frac{N\sqrt{p_1p_2p_3}}{\kappa\ \gamma^2\lambda^2}\Big[\sin(\mub_1c_1)\sin(\mub_2c_2)+\sin(\mub_2c_2)\sin(\mub_3c_3)+\sin(\mub_3c_3)\sin(\mub_1c_1)\Big].
\ee
In the gravitational sector, all the information is contained in the $h_i$:
$
h_1=\mub_1c_1=\lambda\sqrt{\frac{p_1}{p_2p_3}}c_1
$
and cyclic expressions.
By defining the quantum shear by
\be
\sQ:=\frac{1}{3\gamma^2\lambda^2}\left(1-\frac{1}{3}\Big[\cos(h_1-h_2)+\cos(h_2-h_3)+\cos(h_3-h_1) \Big]\right),
\label{sQ}
\ee
one can show \cite{linda2} that LQC-modified generalized Friedman equation is: 
$
H^2=\sQ+\frac{\kappa}{3}\rho-\lambda^2\gamma^2\left(\frac{3}{2}\sQ+\frac{\kappa}{3}\rho\right)^2.
\label{fried}
$

In \cite{linda3}, exhaustive numerical simulations to investigate the duration of inflation as a function of the different variables entering the dynamics in Bianchi-I LQC were carried out. 
As the shear is initially small compared to everything else, the initial conditions for the matter content are chosen \cite{linda1} as
$
\rho(0)=\rho_0\left(1-\frac{1}{2}\sqrt{3\kappa\rho_0}\frac{1}{2m}\sin(2\delta)\right)^{-2},
$
$
m\phi(0)=\sqrt{2\rho(0)}\sin(\delta),
$
and
$
\dot\phi(0)=\sqrt{2\rho(0)}\cos(\delta),
$
where $\rho_0$ is the initial energy density up to a small correction, and $\delta$ is the phase of the oscillations between the kinetic and the potential energy. 
The phase and shear are the initial variables to set.

\begin{figure}
	\begin{center}
		\includegraphics{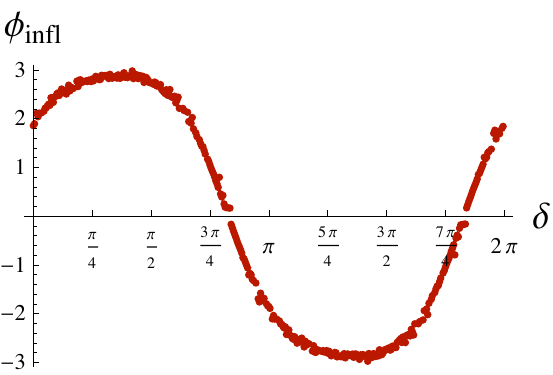} \hfill 
		\includegraphics{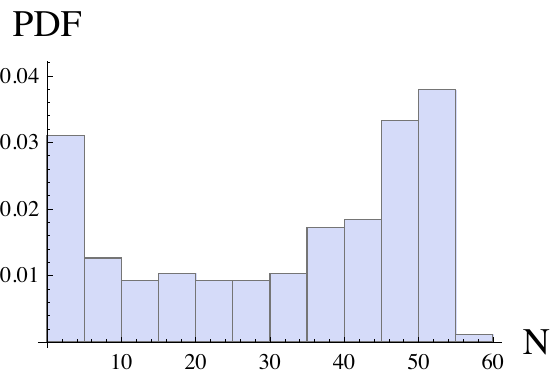}\\
		\includegraphics{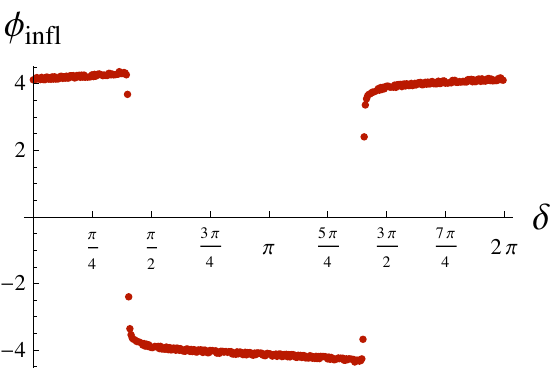}\hfill 
		\includegraphics{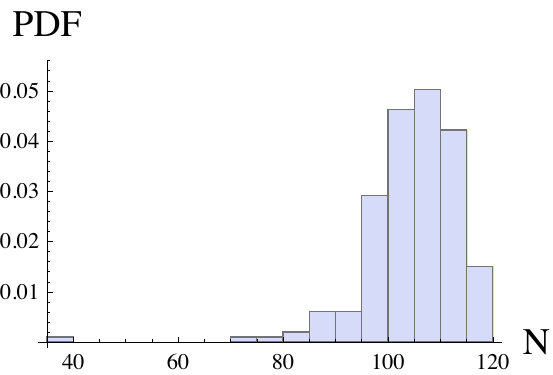}\\
		\caption{Results of the simulations carried out in  \cite{linda3}. From top to bottom : $\sigma_Q(0)=10^{-2}\frac{\kappa}{3}\rho_0$ and $\sigma_Q(0)=10^{-6}\frac{\kappa}{3}\rho_0$. The first column is $\phi$ at the start of slow-roll inflation and the second column corresponds to the numerically calculated probability distribution function of the number of e-folds of inflation.}
		\label{Big}
	\end{center}
\end{figure}

Some results of the simulations are showed in Fig. \ref{Big}. The main conclusion is that, in general, the number of e-folds decreases when the shear increases. But a greater shear will also lead to a larger spread in the number of e-folds, depending on the initial angle $\delta$. The number of e-folds of slow-roll inflation depends strongly on  $\rho_{max}$ which is fixed only when the shear vanishes. 
At the bounce, the dynamics is completely driven by the kinetic energy and the shear. The kinetic energy grows a lot in a very short time, which gives the scalar field a boost, and lifts it up to create the initial conditions for slow-roll inflation. If the shear is important, the bounce will happen at a lower value of the kinetic energy, and the scalar field potential will not be "climbed" as high as in the isotropic case.

Anisotropies lead to fewer e-folds of slow-roll inflation. It is however interesting that for a wide range of parameters,
the probability distribution  for the number of e-folds is peaked at values compatible with data, between 70 and 130 e-folds. It is worth noticing that whereas any value between 0 and $N_{max} = 2\pi \sqrt{2\rho_c} m^{-2} = 3.9\times 10^{12}$ is {\it a priori} possible for $N$, the favored value is very close to the minimum required value. This makes the bounce/inflation scenario particularly appealing for phenomenology: all the quantum information from the bounce might not have been  washed out by inflation. Having $N$ close to 70 is exactly what is required to lead to measurable effects in the CMB spectrum. An important issue however remains: what would be a "natural" initial value for the shear?

\section{Cosmology : direct probes}\label{ra_sec3}
Directly probing LQC modeling of the universe from astronomical observations follows the standard procedure used in classical cosmology to probe {\it e.g.} the physics of inflation. Any observer is confined within the Universe and one relies on cosmic inhomogeneities (whose evolution across cosmic times depends on the dynamics of the Universe) as internal tracers. They are revealed by the observed galaxies and large scale structures, and by the anisotropies of the cosmic microwave background (CMB). This however tells us that our Universe is {\it statistically} homogenous and isotropic, being filled with inhomogeneities, and can be modeled by a {\it perturbed} FLRW metric, for those inhomogeneities are small in the primordial Universe.

In classical cosmology, inhomogeneities are produced during inflation from the gravitational amplification of the fluctuations of the quantum vacuum. In the context of single field inflation, the perturbations are of two types: scalar modes corresponding to perturbations of the scalar 3-dimensional curvature, denoted $\mathcal{R}$, and, tensor modes $h^i_a$ corresponding to primordial gravitational waves. They are commonly described by the Mukhanov-Sasaki, gauge-invariant variables, $v_\mathrm{S}=z\mathcal{R}$ and $v_\mathrm{T}=ah$ with $z=(a\bar\varphi')/\mathcal{H}$ and $\mathcal{H}=a'/a$ is the Hubble parameter in conformal time, {\it i.e.} $ds^2=a^2(\eta)\left(d\eta^2-dx_adx^a\right)$. The quantum fluctuations of these two fields are dynamically amplified during the accelerated expanding inflationary phase. Since they originate from the quantum vacuum which is Gaussian (and assuming linear evolution for simplicity), the perturbations at the end of inflation, $\eta_e$, are fully described by their 2-points correlation function or, in Fourier space, by their primordial power spectrum:
\begin{eqnarray}
	\mathcal{P}_\mathrm{S}=\frac{k^3}{2\pi}\left<\mathcal{R}(k)~\mathcal{R}^\star(k)\right>_{\eta_e},&\mathrm{and}&\mathcal{P}_\mathrm{T}=(16Gk^3)\displaystyle\sum_{s=1}^2\left<h^i_{a,(s)}(k)~h^a_{i,(s)}(k)\right>_{\eta_e},
\end{eqnarray}	
where the average is a quantum expectation value over the vacuum state. For tensor modes, the sum is over the two helicity degrees of freedom. At the end of inflation, our Universe is then filled with inhomogeneities of quantum origin: scalar perturbations serves as the primordial seeds for structures formation, and, both scalar {\it and} tensor perturbations leave their footprint in the CMB in the form of anisotropies of temperature and linear polarization. The latter is decomposed into two modes dubbed $E$ and $B$ modes. The statistics of these anisotropies follows the statistics of the cosmological perturbations and is gaussian (primordial non-gaussianities are observationally constrained to be extremely small). The observed information contained in the CMB is then compressed into six angular power spectra measuring the power of the $T,~E$ and $B$ auto- and cross-correlations. These are estimated from the CMB observations and are theoretically related to the primordial power spectra via the line-of-sight solution of the Boltzmann equations \cite{los}:
\begin{equation}
C^{XY}_\ell=\displaystyle\int^\infty_0 dk\int^{\eta_0}_{\eta_e}d\eta \left[\Delta^{X,\mathrm{S}}_\ell(k,\eta)
\Delta^{Y,\mathrm{S}}_\ell(k,\eta)\mathcal{P}_\mathrm{S}(k)+\Delta^{X,\mathrm{T}}_\ell(k,\eta)\Delta^{Y,\mathrm{T}}_\ell(k,\eta)
\mathcal{P}_\mathrm{T}(k)\right], \label{eq:los}
\end{equation}
with $X,~Y$ running over $T,~E$ and $B$. The time integration is performed from the end of inflation to today, $\eta_0$. The functions $\Delta^{X,\mathrm{S(T)}}_\ell$ are the transfer functions encoding the evolution of scalar(tensor) perturbations and the primordial power spectra are source terms. Fitting the predicted angular power spectra on the estimated ones allows for setting contraints on both cosmological parameters driving the dynamics of the homogeneous Universe via the transfer functions and cosmological parameters driving the shape of the primordial power spectra. Since the later are classically derived from the inflationary dynamics, any constraints on $\mathcal{P}_\mathrm{S(T)}$ from the CMB measurements can be translated into constraints on inflationary models.

In the context of LQC, the cosmological perturbations evolve through the contracting phase and the bounce prior to inflation. Because of that, one can expect some distortions in the predicted $\mathcal{P}_\mathrm{S(T)}$ as compared to the standard prediction of pure inflation. The shape of primordial power spectra now contains informations about the contracting phase and the quantum bounce in addition to informations about inflation, and this will inevitably translate into distortions of the angular power spectra of the CMB anisotropies, leading to possible direct probes of this quantum gravity modeling of the Universe.  The main prediction is therefore the primordial power spectra from which CMB angular power spectra are derived. Preliminary results were obtained by solely considering the change in the background evolution, the Universe passing through a contraction phase and bounce prior to inflation \cite{prelim}. The distortions on the polarized CMB anisotropies could be observed from a clear inspection of those anisotropies and used to constrain {\it e.g.} the fraction of potential energy in the scalar field at the time of the bounce \cite{obsbounce}. However, the very fact that cosmological perturbations are to be constructed from a quantum theory of gravity was not properly taken into account, though the change of the Universe history was. Indeed, cosmological perturbations are perturbations of the gravitational field itself (as well as perturbations of the matter content). This means that the classical theory of cosmological perturbations (consisting in linearizing the Einstein's field equations around the FLRW solution) should be amended first for accounting that perturbations live in a {\it quantum} background.

\subsection{Cosmological perturbations in LQC}
Different approaches to treat cosmological perturbations in a LQC-derived cosmological background have been developed recently. The dressed metric approach, discussed in the chapter by Agulloa and Singh, adopts a minisuperspace strategy in which the homogeneous and isotropic degrees of freedom {\it and} the inhomogoneous degrees of freedom (considered as perturbations) are quantized \cite{aan}. The former is obtained by the loop quantization and the latter is obtained from a Fock quantization on a {\it quantum} background. The physical inhomogeneous degrees of freedom are given by the Mukhanov-Sasaki variables derived from the linearized classical constraints. The second order Hamiltonian (restricted to the square of the first order perturbations) is promoted to be an operator and the quantization is performed using techniques suitable for the quantization of a test field evolving in a quantum background \cite{asht_lew_2009}. The Hilbert space is a tensor product $\Psi(\nu,v_\mathrm{S(T)},\varphi)=\Psi_\mathrm{FLRW}(\nu,\bar\varphi)\otimes\Psi_\mathrm{pert}(v_\mathrm{S},v_\mathrm{T},\bar\varphi)$ with $\nu$ the homogeneous and isotropic degrees of freedom and $v_\mathrm{S(T)}$ the degrees of freedom for perturbations. In the interaction picture, so long as the backreaction of the perturbations on $\Psi_{\rm FLRW}$ remains negligible, the Schr\"odinger equation for the perturbations is shown to be identical to the Schr\"odinger equation for the quantized perturbations evolving in a classical background but using a {\it dressed} metric encoding the quantum nature of the background (for tensor modes):
\begin{equation}
	i\hbar\partial_{\bar\varphi}\Psi_\mathrm{pert}=\frac{1}{2}\displaystyle\int \frac{d^3k}{(2\pi)^3}\left\{\frac{32\pi G}{\tilde{p}_{\varphi}}\left|\hat\pi_{\mathrm{T},\vec{k}}\right|^2\Psi_\mathrm{pert}+\frac{k^2}{32\pi G}\frac{\tilde{a}^4({\bar\varphi})}{\tilde{p}_{\varphi}}\left|\hat{v}_{\mathrm{T},\vec{k}}\right|^2\Psi_\mathrm{pert}\right\},
\end{equation}
with
\begin{eqnarray}
	(\tilde{p}_{\varphi})^{-1}=\left<\hat{H}^{-1}_\mathrm{FLRW}\right> &~\mathrm{and}~&\tilde{a}^4=\frac{\left<\hat{H}^{-1/2}_\mathrm{FLRW}\hat{a}^4({\bar\varphi})\hat{H}^{-1/2}_\mathrm{FLRW}\right>}{\left<\hat{H}^{-1}_\mathrm{FLRW}\right>}.
\end{eqnarray}
In the above, $(\hat{v}_{\mathrm{T},\vec{k}},\hat\pi_{\mathrm{T},\vec{k}})$ are the configuration and momentum operators of the perturbations while $\hat{H}_\mathrm{FLRW}$ is the Hamiltonian operator of the isotropic and homogeneous background. The dressed metric is in principle {\it neither} equal to the classical metric {\it nor} equal to the metric traced by the peak of the sharply peaked background state. This is finally translated into a Fock quantization for which the mode functions (providing the evolution of scalar and tensor perturbations in a quantum background, here expressed in the spatial Fourier space) are solutions of 
\begin{eqnarray}
&&Q''_k+2\left(\frac{\tilde{a}'}{\tilde{a}}\right)Q'_k+\left(k^2+\tilde{U}\right)Q_k=0, \\
&&h''_k+2\left(\frac{\tilde{a}'}{\tilde{a}}\right)h'_k+k^2h_k=0.
\end{eqnarray}
The gauge-invariant variable $Q_k$ is related to the Mukhanov-Sasaki variables for scalar modes via 
$Q_k=v_{\mathrm{S},k}/a$, and, $\tilde{U}$ is a dressed potential-like term given by
\begin{equation}
\tilde{U}({\bar\varphi})=\frac{\left<\hat{H}^{-1/2}_\mathrm{FLRW}\hat{a}^2({\bar\varphi})\hat{U}({\bar\varphi})\hat{a}^2({\bar\varphi})
\hat{H}^{-1/2}_\mathrm{FLRW}\right>}{\left<\hat{H}^{-1/2}_\mathrm{FLRW}\hat{a}^4({\bar\varphi})\hat{H}^{-1/2}_\mathrm{FLRW}\right>},
\end{equation}
the quantum counterpart of
\begin{equation}
U({\bar\varphi})=a^2\left(fV({\bar\varphi})-2\sqrt{f}\partial_{\bar\varphi} V+\partial^2_{\bar\varphi} V\right),
\end{equation}
with $f=24\pi G (\dot{\bar\varphi}^2/\rho)$, the fraction of kinetic energy in the scalar field.\\

A second approach developed in Refs. \cite{jakub_tom} consists in perturbing the semi-classical, effective space-time whose dynamics is given by the modified Friedmann equations. The idea is to start from the classical perturbed Hamiltonian and to introduce corrections taking into account at the effective level the quantum nature of the background. For the zeroth-order Hamiltonian, providing the dynamics of the background, such a modification is easily obtained from the fact that the quantization being based on holonomies, the connection $\bar{k}$ is replaced by $\left(\sin(\gamma\bar\mu\bar{k})/\gamma\bar\mu\right)$, yielding the modified Friedmann equations. Similar effective modifications are introduced to the first and second order perturbation Hamiltonians. Though there is a priori much more freedom for those modifications, there expressions are univocally derived by requiring that first, the classical Hamiltonian is recovered in the limit of large volumes ({\it i.e.} $\bar\mu\to0$), and, second, that the algebra of the truncated scalar, diffeomorphism and Gauss constraints is still closed, as is the case for truncated contraints in the classical theory of cosmological perturbations. This second requirement fixes all the ambiguities of the introduced quantum corrections (at least for the case of holonomy corrections). Moreover, the set of effective contraints is first class and can be used to generate the gauge transformations to derive the effective gauge-invariant variables for the cosmological perturbations. There dynamics is generated by the second-order, effective Hamiltonian. Those perturbations are finally quantized {\it \`a la} Fock using the techniques developed for quantum fields in curved spaces. In that process, it appears that the anomaly-free algebra of effective constrained is deformed compared to the classical algebra of constraints by \cite{tom}: 
\begin{eqnarray}
	\left\{D[M^a],D [N^a]\right\} &=& D[M^b\partial_b N^a-N^b\partial_b M^a],  \\
	\left\{D[M^a],S^Q[N]\right\} &=& S^Q[M^a\partial_a N-N\partial_a M^a],  \\
	\left\{S^Q[M],S^Q[N]\right\} &=& D\left[\mathbf{\Omega} q^{ab}(M\partial_bN-N\partial_bM)\right],
\end{eqnarray} 
with $D$ the diffeomorphism constraint and $S^Q$ the scalar constraint. The deformation is encoded in $\mathbf{\Omega}$ which depends on the background phase-space variables, $\mathbf{\Omega}=\cos(2\gamma\bar\mu\bar{k})=1-2\rho/\rho_c$. In this deformed algebra approach, the mode functions describing the dynamics of the scalar and tensor modes (in terms of {\it effective} Mukhanov-Sasaki variables) are solutions of 
\begin{equation}
	v''_{\mathrm{S(T)},k}+\left[\mathbf{\Omega} k^2-\frac{z''_{\mathrm{S(T)}}}{z_{\mathrm{S(T)}}}\right]v_{\mathrm{S(T)},k}=0,
\end{equation}
with $z_\mathrm{S}=(a\bar\varphi')/\mathcal{H}$ and $z_\mathrm{T}={a}/{\sqrt{\mathbf{\Omega}}}$.
Those functions encodes the impact of the effective background on the perturbations.	

\subsection{Primordial power spectrum in loop quantum cosmology}
\begin{figure}
\begin{center}
	\includegraphics[scale=0.5]{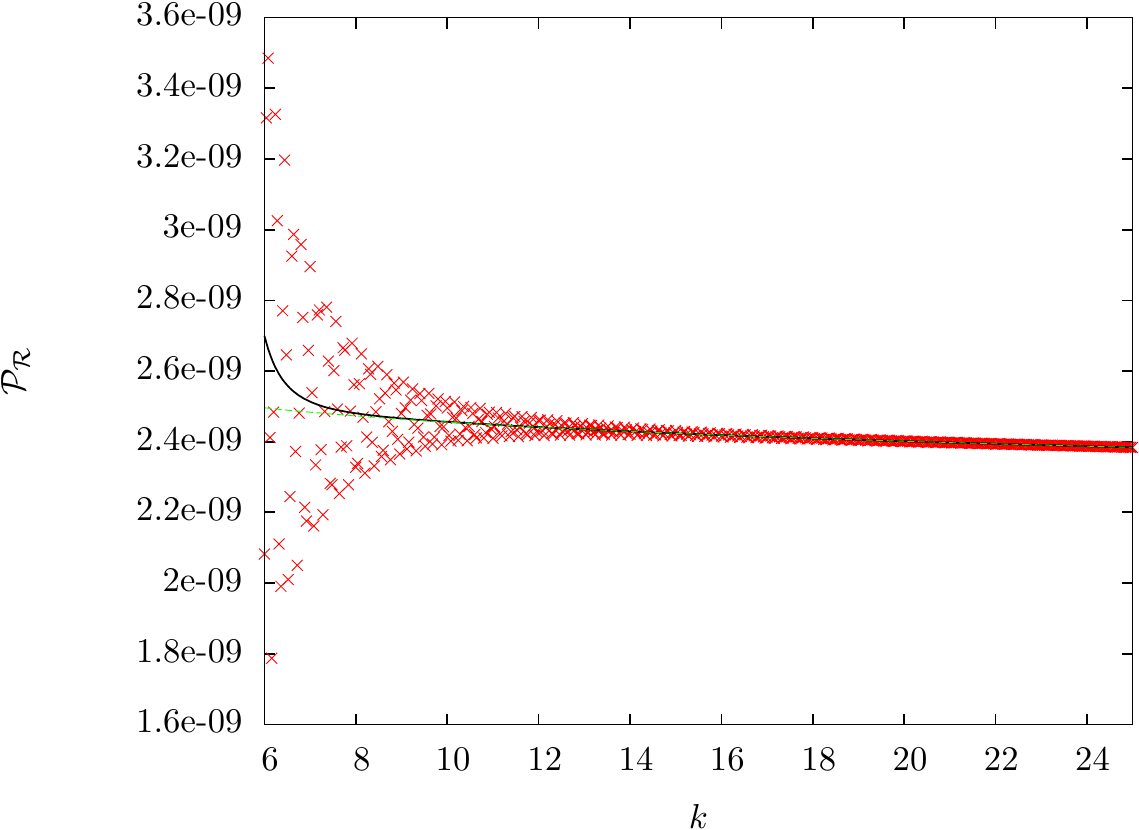}\hspace{0.7cm} \includegraphics[scale=0.5]{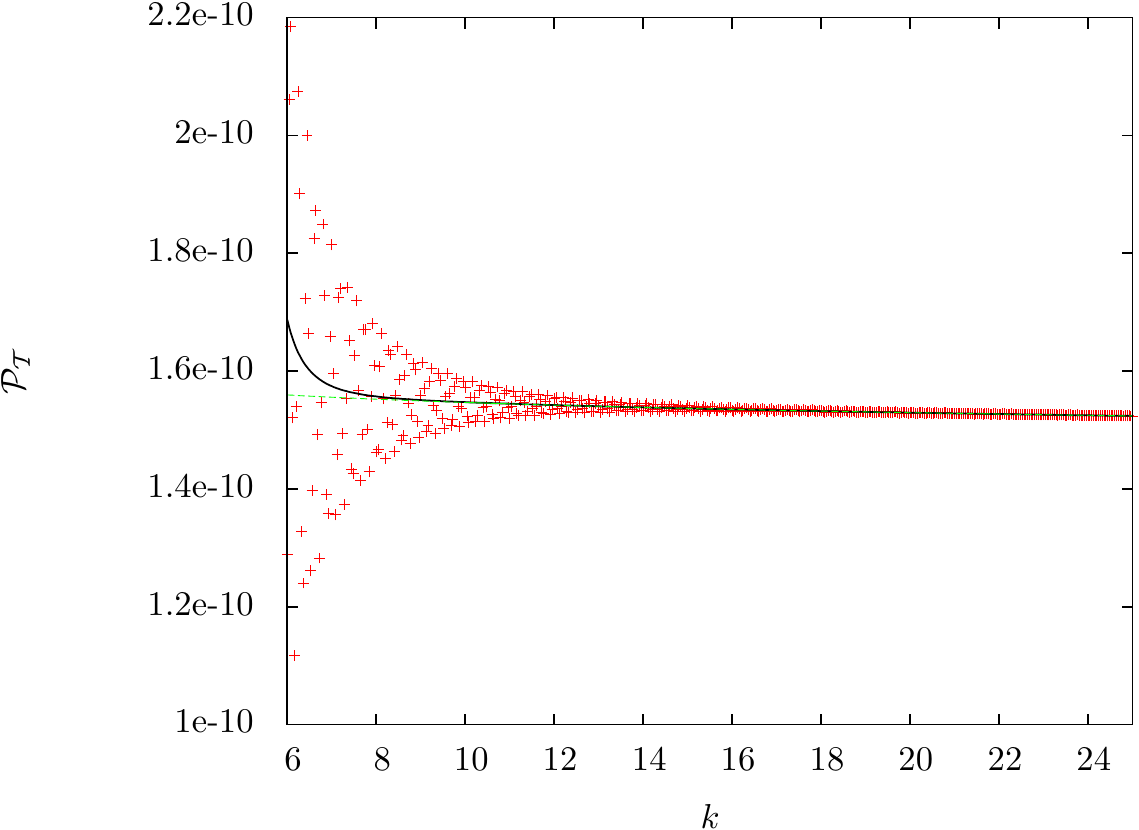}
	\caption{Primordial power spectra for scalar (left) and tensor (right) modes in the dressed-metric approach. Initial conditions are set at the time of the bounce (from \cite{aan}).}
	\label{fig:pkaan}
\end{center}
\end{figure}

The primordial power spectra are the sources of the CMB anisotropies and are the key quantities to compute. Assuming some initial conditions for the mode functions, thus fixing the choice of the initial quantum states for perturbations, the primordial power spectra are determined by the knowledge of the mode functions at the end of inflation. A first choice of initial conditions for perturbations is a fourth order WKB vacuum at the time of the bounce. Such a choice is however only possible in the dressed metric approach. For the deformed algebra, $\mathbf{\Omega}$ is negatively valued at the time of the bounce which prevents the existence of standard oscillatory solutions for the mode functions. An example of the resulting primordial power spectra for scalar and tensor perturbations in the dressed metric approach and setting the initial conditions for perturbations at the time of the bounce is displayed on Fig. \ref{fig:pkaan}. This shows that the bounce leaves a characteristic length scale $(k_\star)^{-1}$ as a typical footprint. For shorter length scales, $k>k_\star$, the predicted primordial power spectrum co\"\i{n}cides with the prediction of standard inflationary cosmology since the slightly red-tilted power law is recovered. However for larger length scales, LQC predicts a different power spectrum (which can be viewed as a running of the spectral index in the language of inflation). This typical scale can be intuitively understood for tensor modes by a clear inspection of $(\tilde{a}''/\tilde{a})$, tracing the effective "curvature" of the background. For sharply peaked states, the dressed scale factor $\tilde{a}$ is very well approximated by the scale factor traced by the peak of the background quantum states, $a$, which is solution of the modified Friedmann equations. At the time of the bounce, $a''/a=8\pi G \rho_c$ and rapidly decreases in the beginning of the expansion. Then, this quantity rapidly increases once the Universe enters its inflationary phase. The shape of the primordial power spectrum is driven by $(k^2-a''/a)$: if $k^2> a''/a$, the modes are oscillatory whereas in the opposite case, the mode functions are a linear combination of growing and decreasing modes. As a consequence, modes at very short scales, $k\gg k_\star$ with $k_\star=\sqrt{8\pi G \rho_c}$, are affected by the background "curvature" during inflation only, explaining why the standard power law is recovered for the primordial power spectrum at these scales. However, the dynamics of modes such that $k\sim k_\star$ is also affected by the background "curvature" at the time of the bounce and one should expect for those modes a discrepancy as compared to the standard prediction of inflation. 

Such a length scale translates into a characteristic {\it angular} scale in the CMB angular power spectra. By denoting $k_H(t_0)=2.3\times10^{-4}$Mpc$^{-1}$ the wavenumber corresponding to the Hubble distance {\it today}, the characteristic angular scales is given, in terms of multipole $\ell\sim1/\theta$, by $\ell_\star\approx k_\star(t_0)/k_H(t_0)$. This angular scale lies in the range of scales observed in the CMB anisotropies if $k_\star(t_0)>k_H(t_0)$. The characteristic length scale $k_\star$ is set at the time of the bounce and is inevitably stretched by the following cosmic expansion leading to $k_\star(t_0)=\sqrt{8\pi G\rho_c}\times e^{-N}$ with $N$ the number of e-folds from the bounce to today. From the fact that $k_\star$ is of the order of the inverse of the Planck length at the time of the bounce and from the knowledge of the number of e-folds from the {\it end} of inflation to today, this scale set by the bounce enters in the observable range if the number of e-folds during inflation is smaller than $\sim90$. If such a characteristic length scale is indeed in the range observed with the CMB, the slight boost of power for $k\lesssim k_\star$ will translate into a slight boost of the angular power spectrum of the CMB anisotropies (as compared to the inflationary prediction) for angular scales $\ell\lesssim\ell_\star$.

\begin{figure}
\begin{center}
	\includegraphics[scale=0.9]{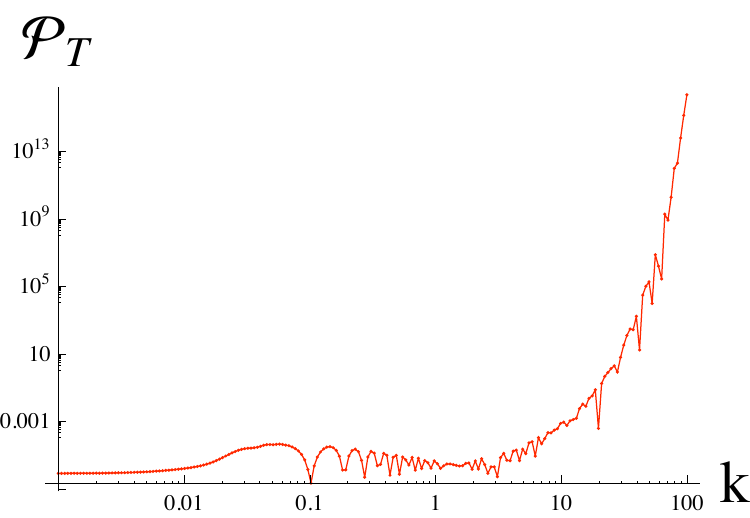}
	\caption{Primordial power spectrum for tensor modes in the deformed-algebra approach (from \cite{linda_pk}). The exponential increase is not necessarily a problem as (i) the observational window might fall out of this region, (ii) the spectrum has anyway a natural cutoff in the UV as the small-scale physics is not described by the primordial spectrum, (iii) backreaction should be taken into account when the amplitude becomes high. The spectrum for scalar modes was derived in \cite{sus}.}
	\label{fig:pknous}
\end{center}
\end{figure}

Another possibility is to set the initial conditions for perturbations deep in the contracting phase. Then, for both the deformed algebra and dressed metric approaches, one can choose a Minkowski vacuum state for all the wavenumbers, $v_{\mathrm{S(T)},k}(\eta\to-\infty)=\exp(ik\eta)/\sqrt{2k}$. In the dressed metric, the standard power law spectrum is recovered for $k\gg k_\star$ from the very same reason as described above: the modes are not affected by the background "curvature" during both the classical contraction and the quantum bounce. In the infrared limit, $k\to0$, the modes are mainly affected by the background during contraction leading to a scale invariant power spectrum. In between, there is a range of modes which are not affected by contraction but by the bounce. In that range of wavenumbers, the primordial power spectrum exhibits oscillations with an envelope exhibiting a boost of the power. As shown in Fig. \ref{fig:pknous}, the prediction differs in the deformed algebra approach \cite{linda_pk}. For modes such that $k>k_\star$, the shape of the primordial spectrum is mainly driven by $\mathbf{\Omega}k^2$. Since $\mathbf{\Omega}$ is negative around the bounce, this leads to an exponential increase of the primordial power spectrum at short scales roughly given by $\mathcal{P}_\mathrm{T}(k\gg k_\star)\propto \exp\left(k\int^{\eta_+}_{\eta_-}\sqrt{\left|\mathbf{\Omega}\right|}d\eta_1\right)$ with $\eta_\pm$ defining the time laps around the bounce during which $\mathbf{\Omega}$ is negative. For larger length scales, $k<k_\star$, the term $\mathbf{\Omega}k^2$ becomes subdominant in the differential equation satisfied by the Mukhanov-Sasaki variable. This regime is therefore very similar to the dressed metric approach previously discussed and the scale invariant behavior in the infrared limit as well as the oscillations for intermediate scales are recovered. A detailed comparisons of both approches was made in \cite{boris}.

Similar studies have been performed for the case of inverse volume (IV) corrections. This includes the derivation of an anomaly-free perturbation theory with IV corrections alone, and, with both holonomy and IV corrections \cite{iv_corrected}. However, the impact of the IV corrections on the bounce itself is not well understood and the primordial power spectra with such corrections has been computed during inflation only. Fortunately, an imprint appears on the largest scales for scalar and tensor modes in the form of a polynomial boost below the pivot scale $k_0$, $\mathcal{P}^{\mathrm{IV}}(k)=\mathcal{P}^{\mathrm{STD}}(k)\times(1+\Gamma\delta_0(k/k_0)^{-\left|\sigma\right|})$ \cite{iv_pk}. Starting from such a predicted power spectrum, the IV parameters have been constrained using WMAP data on the CMB anisotropies showing that {\it e.g.} for $\sigma=2$, the parameter $\delta_0$ is constrained to be smaller than $6.5\times10^{-5}$ at 95\% of confidence level \cite{iv_const}.

\subsection{Measuring the Barbero-Immirzi parameter}
The above results are based on loop quantum cosmology with a {\it real-valued} Barbero-Immirzi parameter, inherited from the standard formulation of loop quantum gravity. Originally, the Ashtekar formulation of gravity as a gauge theory was however built with a complex-valued Barbero-Immirzi parameter, $\gamma=\pm i$, thus simplifying the constraints into being polynomials in the phase-space variables. Though $\gamma$ plays no role at the classical level, it is of primary importance at the quantum level: $\gamma=\pm i$ makes the gauge group to be complex, rendering the quantization difficult. Quantization is usually performed with $\gamma\in\mathbb{R}$ for the gauge group is $SU(2)$, which is directly related to the discreteness of the spectra of geometric operators. The role of $\gamma$ is then crucial in LQC since the discreteness of geometric operators plays an important role in the bounce scenario via the minimal area gap. Phenomenologically speaking, the value of $\gamma$ fixes the value of $\rho_c$ which could be measured by searching for the characteristic scale $k_\star=\sqrt{8\pi G \rho_c}$ in the CMB anisotropies. It was however argued that in the context of three-dimensional gravity, a natural choice would be $\gamma=\pm i$ which still leads to a consistent quantum theory \cite{noui}. This still has to be fully extended to four-dimensional gravity, but this shows that trying to experimentally probe the nature of the Barbero-Immirzi parameter is important.

The two (independent) helicity states of primordial gravitational waves are classically derived from a linearization of Einstein's equations around the inflationary background and subsequently quantized using a Fock scheme on curved spaces. The resulting primordial power spectra for the right-handed and left-handed gravitons are equal, $\mathcal{P}_{r/l}\propto(H/M_\mathrm{Pl})^2$ with $H$ the Hubble parameter during inflation ("graviton" is used to denote a {\it Fock} quantization of tensor modes). The CMB angular power spectrum of the $BB$ correlation is sourced by the sum of the two helicity states ($\mathcal{P}_\mathrm{T}$ in Eq. (\ref{eq:los}) is the sum $\mathcal{P}_{r}+\mathcal{P}_{l}$). The cross-correlations between temperature and $B$-modes (called $TB$), and between $E$- and $B$-modes (called $EB$) are however sourced by the {\it difference} of the two helicity states, $\mathcal{P}_{r}-\mathcal{P}_{l}$. Because $\mathcal{P}_{r}=\mathcal{P}_{l}$ by linearizing Einstein's equations and quantizing {\it \`a la} Fock, $C^{TB(EB)}_\ell$ are vanishing. However, it was argued that primordial gravitons may have a helicity-dependent behavior if linearization is performed in the Ashtekar formalism \cite{chiral}. More precisely, it is argued that if the Barbero-Immirzi parameter is imaginary, the reality condition imposes that at the {\it quantum level}, left-handed and right-handed gravitons do not propagate similarly in an inflationary background, suggesting that linearized gravity may violate parity at the quantum level. (This helicity-dependant behavior only arises if $\gamma$ has an imaginary part and at the quantum level. At the classical level or for a real-valued $\gamma$, there is no such parity breaking in linearized gravity.) If this is indeed the case, the $TB$ and $EB$ angular power spectra are non-zero if $\gamma$ has a non-vanishing imaginary part while these spectra are zero if $\gamma$ is real-valued.

\begin{figure}
\begin{center}
	\includegraphics[scale=0.5]{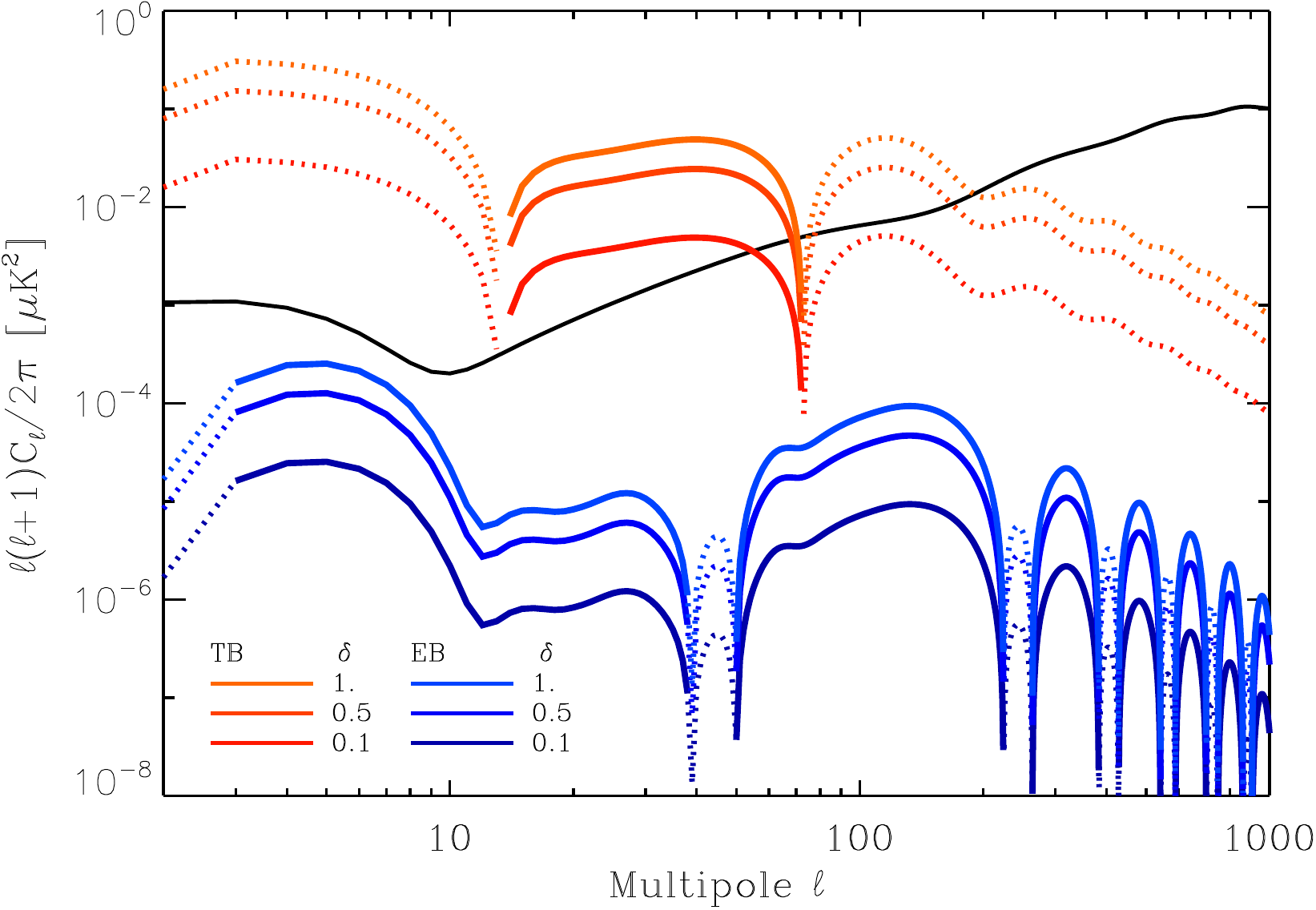}
	\caption{CMB angular power spectra for the $BB$, $TB$ and $EB$ cross-correlations ($r=0.05$) if $\gamma$ is purely imaginary (from \cite{agnes}).}
	\label{fig:tbeb}
\end{center}
\end{figure}

Some $C^{TB}_\ell$ and $C^{EB}_\ell$  (with the $C^{BB}_\ell$ autocorrelation) are depicted in Fig. \ref{fig:tbeb}, including lensing of CMB photons by large scale structures \cite{agnes}. Dotted parts stand for negative values of $TB$ and $EB$ correlations which is an important piece of information since {\it e.g.} a negative $C^{TB}_\ell$ at $\ell\leq15$ corresponds to more power in the right-handed gravitons. The amplitude of the $BB$ autocorrelation is set by the value of the tensor-to-scalar ratio, $r$ (equal to 0.05 in Fig. \ref{fig:tbeb}). Introducing $\delta=\frac{\mathcal{P}_{r}-\mathcal{P}_{l}}{\mathcal{P}_{r}+\mathcal{P}_{l}}$ which amounts the level of parity violation in the linearized gravitational sector, the amplitude of the $TB$ and $EB$ correlations is set by $(r\times \delta)$. A reconstruction of $r$ and $\delta$ is then possible from a measurement of $C^{BB}_\ell,~C^{TB}_\ell$ and $C^{EB}_\ell$. The parameter $\delta$ is a direct measure of the level of parity breaking, and subsequently a direct test of a possible non-vanishing imaginary part of $\gamma$, as $\left|\gamma\right|=\left(1\pm\sqrt{1-\delta^2}\right)/\left|\delta\right|$ for the simplified case of a purely imaginary Barbero-Immirzi parameter.

For a future, highly-sensitive satellite mission dedicated to the CMB polarization, the measurements of polarized $B$-modes would be accurate enough for detecting at least 50\% of parity violation at {\it e.g.} 95\% of Confidence Level (C.L.) for $r=0.2$ (the uncertainties are dominated by the sampling variance). Similarly, measuring $C^{TB(EB)}_\ell$ consistent with zero would lead to an upper bound on $\delta$, directly translated into an exclusion range for $|\gamma|$. For $r=0.05$, the exclusion range at 95\% C.L. is $0.66\leq\left|\gamma\right|\leq1.5$, and it is enlarged to $0.2\leq\left|\gamma\right|\leq4.9$ for $r=0.2$ \cite{agnes}.

\section{Lorentz invariance violation}\label{ra_sec4}
Testing for quantum gravity usually assumes an access to gravitational phenomena for which the curvature becomes close to the Planck scale. In Ref. \cite{amelino}, it was first argued that one can also search for quantum gravity imprints by studying the propagation of particles whose energy is comparable to the quantum gravity energy scale (or even much below if the propagation distance is high enough). The basic idea is that discreteness is a genuine property of the quantum space-time. In the context of LQG, this can be understood from the {\it discreteness} of geometric operators as volume and area operators. This granularity fixes an invariant length scale in apparent contradiction with special relativity (as a boost can contract any length scale). Even though arguments showing that the discreteness of geometric operators can be in agreement with Lorentz invariance have been put forward (see \cite{rovelli_liv}, which argues that the discrete spectrum is observer invariant but the expectation values are not), this granularity idea has opened a wide area of quantum gravity phenomenology aiming at searching for Lorentz invariance violation or deformation as a tracer of quantum gravity. This rich phenomenology is encoded in the fact that the energy-momentum dispersion relation is modified $E\simeq p+m^2/2p\pm\xi(E^2/M_{QG})^n$, with $M_{QG}$ the energy scale of quantum gravity, $\xi>0$, and $n$ usually chosen as an integer. Because of that, the group velocity for {\it e.g.} photons becomes momentum dependent. This means that two photons emitted at the same time but at different momenta would be received at two different times by a distant observer, as, (for $n=1$), $\Delta v\simeq \xi\Delta k D/M_{QG}$ with $\Delta k$ the momentum difference and $D$ the distance from the emitter to the receiver. One should therefore look for  energetic phenomena (thus $\Delta k$ is close enough to the quantum gravity scale) and cosmological distances (for having a cumulative impact) for such an effect to be detectable. 

If Lorentz invariance is indeed broken or deformed by quantum gravity, this could be described at an effective level. There are many different ways of implementing this idea, ranging from non-commutative space-time to effective field theories and non-linear Poincar\'e symmetries. Here, we only mention a few which are closely related to LQG and refer the interested reader to \cite{girelli} and references therein for a detailed presentation. In all the implementations discussed here, one arrives at a modified dispersion similar to the one mentioned above, with a potentially additional helicity dependance. One approach consists in analyzing the Hamitonian of the electromagnetic field in a semi-classical state being an discrete approximation of the flat geometry, dubbed a weave \cite{gambini_liv}. Because the densitized triad operator enters  the Hamiltonian for electromagnetism, its expectation value on the weave state is expected to receive loop quantum gravity corrections. The resulting modified dispersion relation for photons acquires a helicity-dependant correction $\omega^2_{\pm}=k^2\mp4\chi k^3/M_{Pl}$ with $\chi\sim1$. In such a case, photons would experience birefringence in vacuum modifying their polarization state. This effect has been investigated (in the framework of effective field theory though) in \cite{crab} and \cite{gleiser}.

Another approach was put forward in \cite{smolin_liv}. The idea is that, classically, the action functional $S[A]=\int_\Sigma\mathcal{S}[A]$ can be used to define a slicing of the classical space-time. If one now considers a quantum setting, this slicing fluctuates around the classical neighborhood corresponding to space-time variations. The explicit calculation performed in \cite{smolin_liv} considers a Born-Oppenheimer state $\Psi_0[A]\chi[A,\phi]$ with $\Psi_0$ a semiclassical state peaking at the classical solution and $\phi$ a matter field. The expectation value of the densitized triad on such a semiclassical state, evaluated around the classical trajectory, is deformed to $E^{(0)~a}_{~~~i}(x,t,\omega)=E^{(0)~a}_{~~~i}(x,t)(1-\alpha L_{Pl}\omega)$ with $E^{(0)~a}_{~~~i}(x,t)$ the classical solution and $\omega$ to be interpreted as the energy of the matter field (in the sense that $\chi[t,\phi]\propto e^{-i\omega t}\chi_\omega[\phi]$). The time parameter $t$ is defined from the action functional $\mathcal{S}[A]$. Since the triad is now $\omega$-dependent, this defines an $\omega$-dependent metric and thus a modified dispersion relation: $m^2=\omega^2-k^2/(1-\alpha L_{Pl}\omega)$. A potential interpretation is that quantum gravity fluctuations lead to an effective frame in which momenta are measured \cite{liberati}. Classically, the physical momenta $p^a$ is measured in a local inertial frame fixed by the space-time manifold, $p_a=e_a^\mu\pi_\mu$ with $\pi_\mu$ interpreted as the generator of translations. Quantum fluctuations of the space-time itself would then lead to an effective frame $\tilde{e}_a^\mu$ which is non-linearly related to $e_a^\mu$ with a $\pi_\mu$ dependance, $\tilde{e}_a^\mu=F(e_a^\mu,\pi_\mu)$. Since the physical momenta are now measured by $\tilde{p}_a=\tilde{e}_a^\mu\pi_\mu$, the transformation law for momenta would not be given anymore by the Lorentz matrices. In that case, one is therefore considering a deformation of the Poincar\'e symmetry since the relativity principle is preserved but the transformation rules are now non-linear \cite{magueijo_liv}.

\section{Black holes}\label{ra_sec5}

Black holes have been extensively studied in loop quantum gravity (see the chapter by Barbero and Perez in this book). As their macroscopic structure hopefully coincides (up to very small corrections) with the one predicted by general relativity, it is very hard to test LQG with the observation of black holes. Recovering the correct value of the entropy is a very powerful consistency test but can hardly be considered as an experimental confirmation. The only way to observationally investigate LQG with black holes would probably be through their Hawking evaporation. As no evaporating black hole has been seen up to now, this is a prospective work. However, a wide variety of phenomena, reviewed in \cite{carr}, can in principle lead to primordial black holes. 

The idea proposed in  \cite{barrau_pbh} is to search for possible LQG signatures in the spectrum of evaporating black holes. 
The state counting for black holes in LQG relies on the isolated horizon framework (that is a boundary of the underlying manifold considered before quantization). For a given area $A$ of a  black hole horizon, the states arise from a punctured sphere whose punctures carry quantum labels (see, {\it e.g.},
\cite{diaz2}). Two labels $(j,m)$ are assigned to each puncture, $j$ being a
spin half-integer with information on the area and $m$ being its associated projection
with information on the curvature. They satisfy the condition
$
\label{eq1}
A-\Delta\leq 8\pi \gamma \ell_P^2\sum_{p=1}^N{\sqrt{j_p(j_p+1)}}\leq A+\Delta,
$
where $\gamma$ is the  Barbero-Immirzi parameter of LQG, $\Delta$ is a ``smearing'' parameter 
and $p$ labels the different punctures. One may also add the closure constraint:
$
\label{eq2}
\sum_{p}{m_p=0},
$
which corresponds to a horizon  with spherical topology. \\

In the past, it was postulated that due to quantum gravitational effects, the change in the area of a
black hole should be proportional to a fundamental area of the order of the Planck area one. 
It was then hoped that associated lines in the
evaporation spectrum should appear and might reveal quantum gravity effects. However it was understood in \cite{rovelli1, krasnov}
that the situation is  different in LQG because the spacing of the energy 
levels decreases exponentially with the energy. In  \cite{barrau_pbh},  this issue was readdressed and it was shown that several different signatures can in fact be expected.

\begin{figure}[ht]
	\begin{center}
		\includegraphics[scale=0.45]{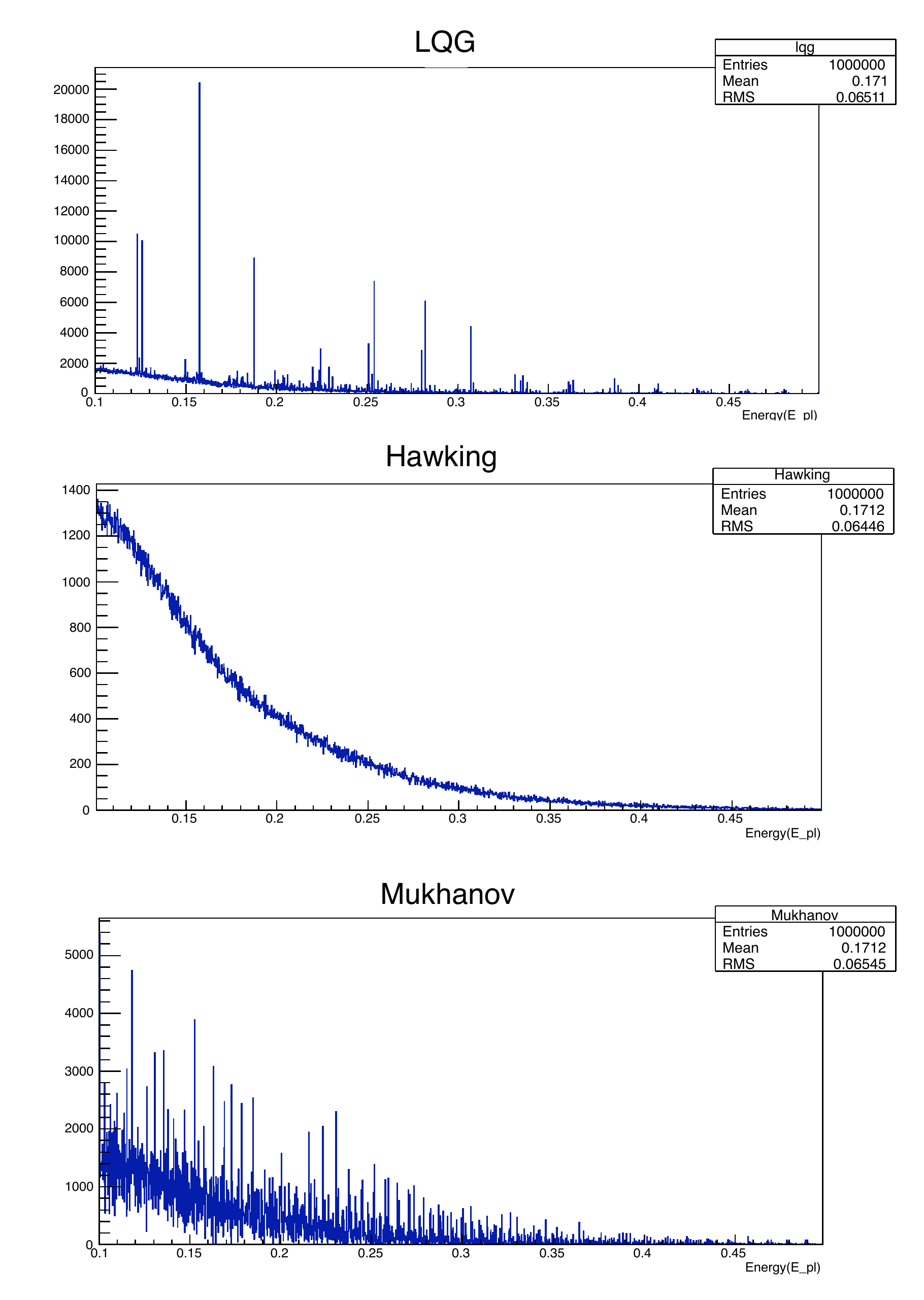}	
		\caption{Spectrum of emitted particles in LQG, in the pure Hawking case, and with an area proportional to the Planck area (Mukhanov), from top to bottom.}
		\label{figevap}
	\end{center}
\end{figure}

To investigate the evaporation in the deep quantum regime, a dedicated and optimized algorithm was developed. It is based on  \cite{diaz1}  and improved by a breadth-first search. 
To see if there is a measurable difference, the evaporation has been considered both according to the pure Hawking law and according to  LQG. In each case, it was modeled by expressing the probability of transition between states as the exponential of the entropy difference modulated by the greybody factor. Those factors were computed beyond the optical limit by solving the quantum wave equations in the curved background of the black hole.  Fig.~\ref{figevap} shows that some specific lines associated with 
transitions occurring during the  last stages of the evaporation can be identified in the LQG
spectrum whereas the pure Hawking spectrum is naturally featureless. 

\begin{figure}[ht]
	\begin{center}
		\includegraphics[scale=0.45]{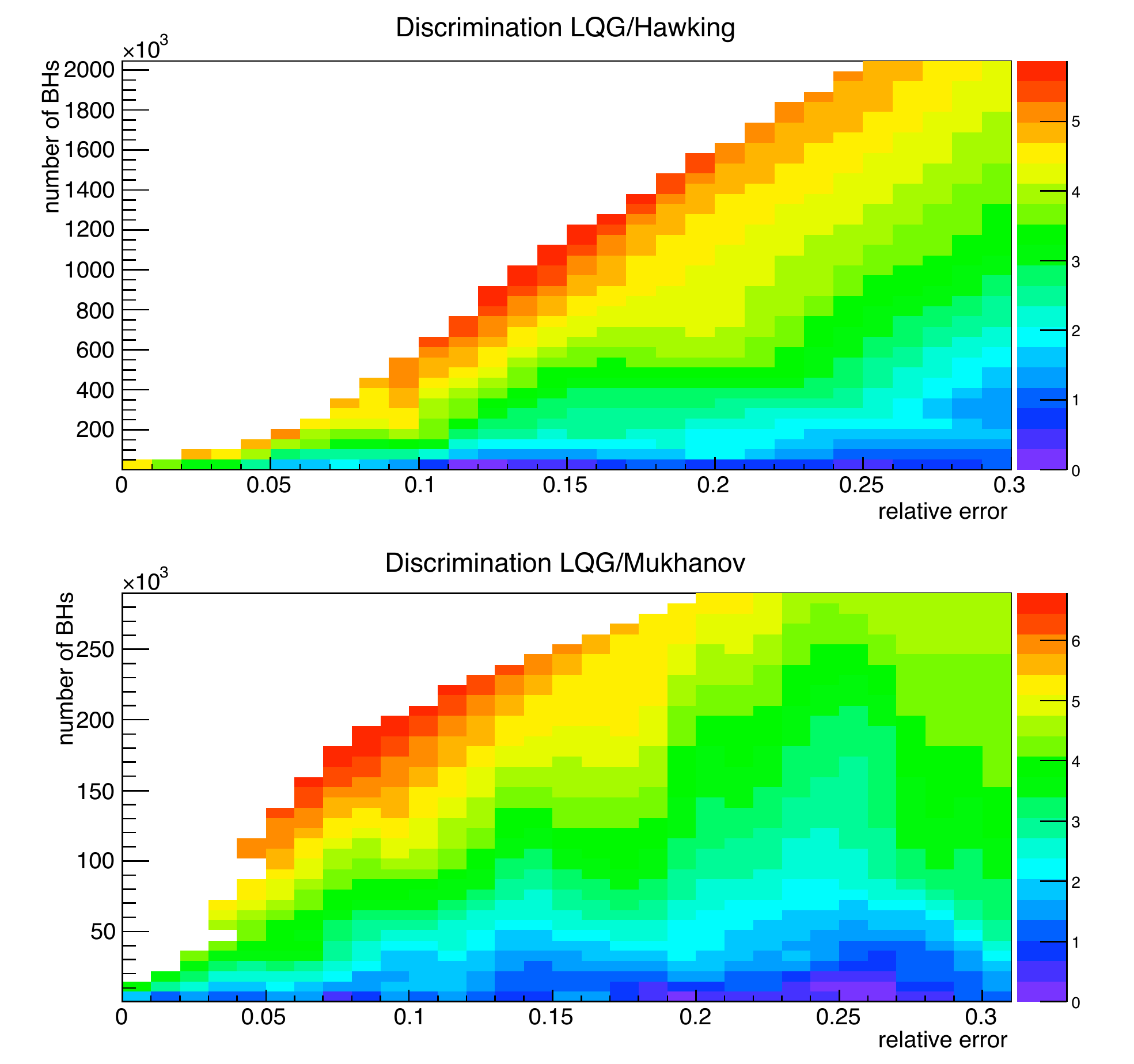}
		\caption{Number of evaporating black holes that should be observed as a function
		of the  error on the energy reconstruction of the emitted leptons for
		different confidence levels (the scale corresponds to the number of standard
		deviations). Up : discrimination between LQG and the
		Hawking hypothesis. Down : discrimination between LQG and 
		the "area proportional to the Planck area" hypothesis.}
		\label{figks}
	\end{center}
\end{figure}

Monte-Carlo simulations were performed to estimate the energy resolution and the number of black holes required for distinguishing between the different scenarios. At each step of the evaporation process, the energy of the
emitted quantum was randomly chosen according to the relevant statistics and to the (spin-dependent)
greybody factor. A Kolmogorov-Smirnov (K-S) test was performed to quantify the distance between
the cumulative distribution functions and used for a systematic study of
possible discriminations between models. Figure~\ref{figks} shows the number of black holes that would have to be
observed for different confidence level in distinguishing between models, as a function of the relative error of the
energy reconstruction. With either enough black holes or a relatively small error, a
discrimination is possible, therefore showing to a clear LQG footprint in the evaporation spectrum. In this study, only emitted leptons were considered to avoid taking into account complicated fragmentation effects. For a detector located close to the black
hole, and due to the huge Lorentz factors, the electrons, muons and taus can be considered as
stable.\\

There is another specific feature of the end-point of the evaporation process which can also be considered. In LQG,
the last transitions take place at definite discrete energies associated with the final peaks in the mass
spectrum whereas in the usual Hawking picture, the simplest way to implement a 
minimal mass is to perform a truncation of the standard spectrum to ensure energy conservation. This leads to the
consequence that in the standard picture, the energy of the emitted quanta will progressively decrease and asymptotically
approach to zero. This ``low-energy'' emission associated with the
end-point can be distinguished from the ``low-energy'' particles emitted earlier in the evaporation process 
thanks to the dynamics. The time interval between consecutive emissions will  
increase with decreasing energies as $E^{-3}$. At 100 TeV, the mean interval is around 1~s.
This specific feature of the ``standard'' spectrum is very different from the absence of 
low-energy particles expected in the LQG case.\\

A final possible test is associated with the pseudo-periodic ``large scale" structure of the area spectrum (see \cite{diaz1} and references therein). Most recent arguments suggest that this periodicity is damped for high masses. If, however, it was to remain, this would lead to interesting features. The  area gap $d A$ between peaks can be shown to be independent of the scale. As, for a Schwarzschild black hole, $dA=32\pi MdM$ and $T=1/(8\pi M)$, this straightforwardly leads to $dM/T=const$ where $dM$ refers to the mass gap between peaks. This is the important point for detection: in units of temperature, the mass gap does {\it not} decrease for increasing masses. Any observable feature associated with this pseudoperiodicity can therefore be searched for through larger black holes. If primordial black holes are formed with a definite mass (as expected for example from phase transitions) and not with a continuous spectrum, their resulting emission can be shown \cite{barrau_pbh} to exhibit potentially detectable features associated with this pseudo-periodicity.\\

A new proposal about statistics, holography, and black hole entropy in loop quantum gravity was suggested in \cite{perez}. The main change is that the degeneracy of area eigenvalues of LQG is now modified in a  simple way by taking into account vacuum fluctuations in the near horizon region. The area spectrum will not be modified but instead of having basically a degeneracy of $2j+1$ for each puncture state, we would now have $e^{a_j/4}$ (where $a_j$ is the area eigenvalue, that is $8\pi \gamma\ell_p^2 \sqrt{j(j+1)}$). Importantly, punctures should in this case be considered as indistinguishable bosons. The very same Monte Carlo simulation approach is being performed to account also for this new model.

\section{Planck stars}\label{ra_sec6}

Recently another idea about black holes and possible observational consequences was pushed in \cite{rovellividotto}. 
The key insight comes first from lessons from quantum cosmology. In loop cosmology, the Friedmann equation is modified by quantum gravitational effects by a term determined by the ratio of $\rho$ to a Planck scale density $
\rho_{\scriptscriptstyle Pl}.
$
The quantum gravity regime seems to be reached when the energy density of matter reaches the Planck scale, $\rho\sim\rho_{\scriptscriptstyle Pl}$. 
The point is that this may happen well before relevant lengths $l$ become planckian.  The bounce is due to a quantum-gravitational repulsion which originates from the Heisenberg uncertainty and does not happen when the universe is of planckian size but instead happens when the  energy density reaches the Planck density. Quantum gravity could become relevant when the volume of the universe is some 75 orders of magnitude larger than the Planck volume \cite{vand}.  

The analogy between quantum gravitational effects on cosmological and black-hole singularities has been successfully used to make a proposal as to how quantum gravity could also resolve the singularity at the center of a collapsed star.  It is assumed that the energy of a collapsing star and any  energy falling into the hole could condense into a highly compressed core with density of the order of the Planck density.  If this is the case, the gravitational collapse of a star does not lead to a singularity but to an additional phase in the life of a star: a quantum gravitational phase where the gravitational attraction is balanced by a quantum pressure. A star in this phase is called a ``Planck star".  The key observation is that a Planck star can have a size
$
   r\sim \left(\frac{m}{m_{\scriptscriptstyle Pl}}\right)^n\  l_{\scriptscriptstyle Pl}
$
where $m$ is now the mass of the star and $n$ is positive.   For instance, if $n=1/3$ (as can be naively computed), a stellar-mass black hole would collapse to a Planck star with a size of the order of $10^{-10}$ cm, that is 30 orders of magnitude larger than the Planck length. The main hypothesis  is that a star so compressed would \emph{not} satisfy the {classical} Einstein equations anymore, even if huge compared to the Planck scale,  because its energy density is already planckian.

The event horizon is replaced by a ``trapping'' horizon \cite{Ashtekar:2005cj} which looks like the standard horizon locally, but from which matter can eventually bounce out.  
The core, that is the ``Planck star", retains memory of the initial collapsed mass $m_i$. In particular, primordial black holes exploding today may produce a distinctive signal for this scenario. Let $m_f = a m_i$ be the final mass reached 
by the black hole before the dissipation of the horizon. It was shown in \cite{rovellividotto}, using  arguments based on information conservation avoiding the firewall hypothesis, that the preferred value is
$
a \sim \frac{1}{\sqrt{2}}.
\label{m}
$
The whole observational scenario relies on the assumption that when the black hole reaches this mass it releases all its energy. \\

During the evaporation phase, the mass loss rate is given by 
$
\frac{{\rm dm}}{{\rm d}t}
=-\frac{f(m)}{m^2},
\label{loss}
$
where $f(m)$ is given above each threshold by  
$
f(m)\approx(7.8\alpha_{s=1/2}+3.1\alpha_{s=1})\times 10^{24}~{\rm g}^3{\rm s}^{-1},
\label{fm}
$
where $\alpha_{s=1/2}$ and $\alpha_{s=1}$ are the number of degrees of freedom (including spin, charge and color) of the emitted particles. If $f(m)$ is assumed to be constant, this leads to:
\be
m_i=\left(\frac{3t_Hf(m_i)}{1-a^3}\right)^\frac{1}{3}.
\label{mi}
\ee
To account for the smooth evolution of $f(m)$  a numerical integration can be carried out and leads to 
$
m_i\approx6.1\times 10^{14}~{\rm g},
\label{minum}
$
and
$
m_f\approx4.3\times 10^{14}~{\rm g}.
\label{mfnum}
$
The value of $m_i$ is very close to the usual value $m_*$  corresponding to black holes needing a Hubble time to fully evaporate. This was expected as the process is explosive. The size of the black hole when it reaches $m_f$ is the only scale in the problem and therefore fixes the energy of the emitted particles in this last stage. All quanta are assumed to be emitted with the same energy taken at 
$
E_{burst}=hc/(2r_f)\approx 3.9 ~{\rm GeV}.
$ \\

Most of the emitted gammas are not emitted with the energy $E_{burst}$ but, instead, come  from the decay of hadrons produced in the jets of quarks. If one assumes that the  branching ratios are controlled by the internal degrees of freedom, the direct emission represents only a small fraction (1/34 of the emitted particles). 
To simulate this process, the "Lund Monte Carlo" PYTHIA code was used to generate the mean spectrum expected for secondary gamma-rays emitted by a Planck star reaching the end of its life. The main point worth noticing is that the mean energy is of the order of $0.03\times E_{burst}$, that is in the tens of MeV range rather than in the GeV range, with a high multiplicity of 10 photons per $q\bar{q}$ jet. 

\begin{figure}
\centerline{\includegraphics[width=9.3cm]{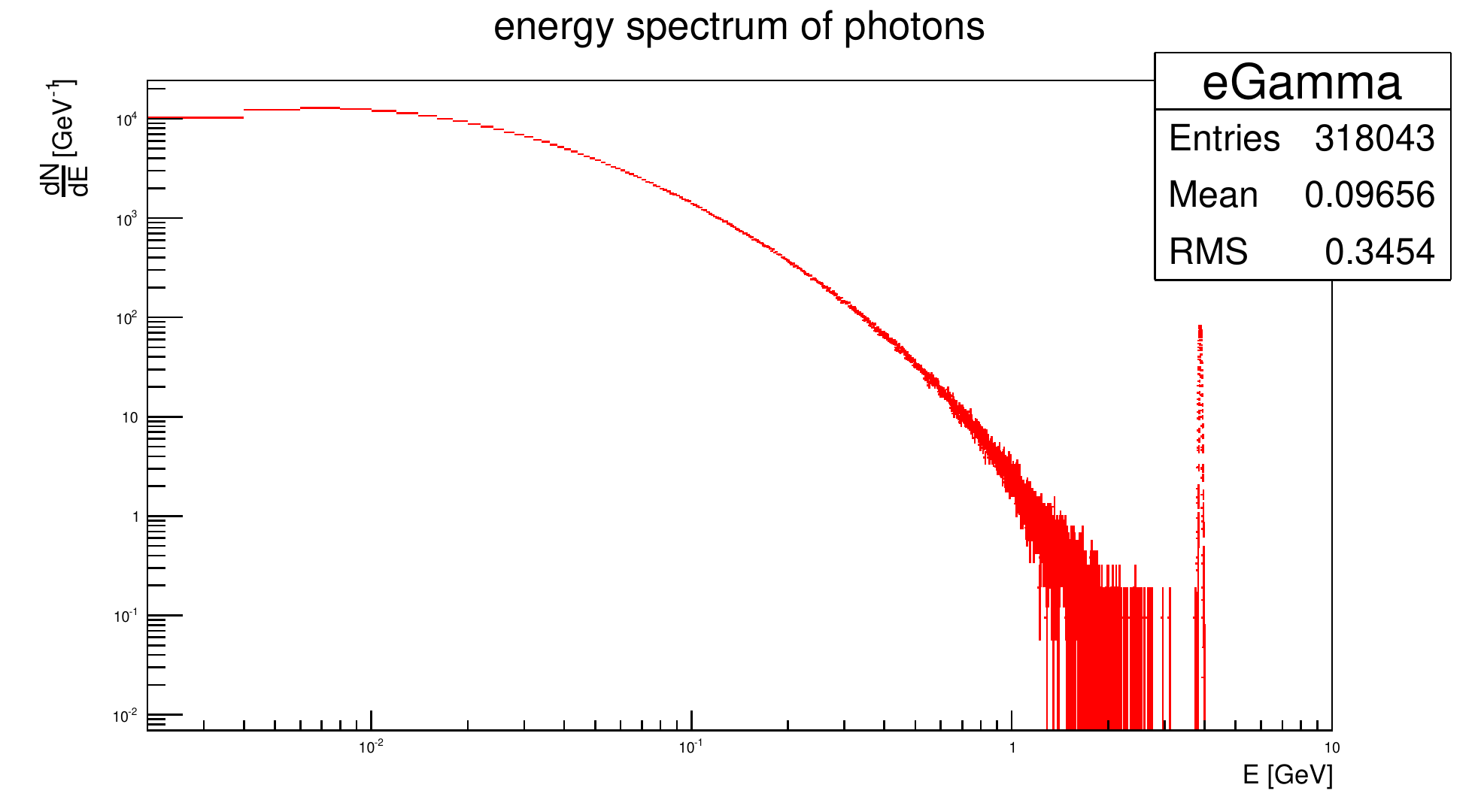}}
\caption{Full spectrum of gamma-rays emitted by a decaying Planck star (log scales).}
\label{full}
\end{figure}

It is straightforward to estimate the number of photons $<N_{burst}>$  emitted during the burst. As for a black hole radiating  by the Hawking mechanism, the particles emitted during the bursts (that is those with $m<E_{burst}$) are emitted proportionally to their number of internal degrees of freedom: gravity is democratic. The spectrum resulting from the emitted $u,d,c,s$ quarks ($t$ and $b$ are too heavy), gluons and photons is shown on Fig. \ref{full}. The little peak on the right corresponds to directly emitted photons that are clearly sub-dominant. By also taking into account the emission of neutrinos and leptons of all three families (leading to virtually no gamma-rays and therefore being here a pure missing energy), one obtains a total number of photons emitted of $<N_{burst}>\approx4.7\times 10^{38}$.

If one assumes a 1 m$^2$ detector, this leads to a maximum distance of detectability of $R\approx 205$ light-years. The ``single event" detection of exploding Planck stars is therefore {\it local} and only a tiny galactic patch around us can be probed. The signal is therefore expected to be {\it isotropic}.

If Planck stars reaching $m_f$ were to saturate the dark matter bound their number within this detectable horizon would be 
\be
N_{det}^{max}=\frac{4\pi \rho^{DM}_*}{3m_f}\left(\frac{S<N_{burst}>}{4\pi N_{mes}}\right)^\frac{3}{2}\approx 3.8 \times 10^{22}.
\label{dist}
\ee
However the usual constraint on primordial black holes $\Omega^{PBH}<10^{-8}$ for initial masses around $10^{15}$ g basically holds and this leads to 
$
N_{det}< 3.8 \times 10^{14},
\label{dist}
$
which is still a high number showing that the individual detection is not impossible.\\

It is possible to estimate the number of events observable in a time  $\Delta t$ corresponding to Planck stars that have masses between $m_f$ and $m(\Delta t)$ at the beginning of the observation time, within the volume $R<R_{det}$. In this case, $m(\Delta t)$ is simply:
$
m(\Delta t)=\left(m_f^3+3f(m)\Delta t\right)^\frac{1}{3}. 
\label{mdeltat}
$
The number of expected "events" during $\Delta t$ is given by
\be
N(\Delta t)=\frac{\int_{m_f}^{m(\Delta t)}\frac{{\rm d}n}{{\rm d}m}{\rm d}m}{\int_{m_f}^{m_{max}}\frac{{\rm d}n}{{\rm d}m}{\rm d}m}\Omega^{PBH}N_{det}^{max}\Omega_{sr},
\label{evts}
\ee
where $m_{max}$ is the maximum mass up to which we assume the mass spectrum 
${\rm d}n/{\rm d}m$ 
to be "filled" by black holes and $\Omega_{sr}$ is the solid angle acceptance of the considered detector. An upper limit on the value of $\Omega^{PBH}$  can be taken conservatively at $10^{-8}$. If one sets $m_{max}=m_*$ and a density of a few percents of the maximum allowed density, that is $\Omega^{PBH}\sim 10^{-10}$, this leads to one event per day.\\

Could such events be associated with some gamma-ray bursts (GRBs) already detected? The long GRBs are well understood and have no link with Planck stars. Were Planck star explosions  to be associated with some of the known GRBs, this would be with short gamma-ray bursts (SGRBs). Interestingly, SGRBs are the less well understood; the redshifts are not measured for a large fraction of them; they are known to have a harder spectrum and some of them do indeed reach the energies estimated here; and a sub-class of SGRB, the very short gamma ray bursts (VSGRBs), do exhibit an even harder spectrum and can be assumed to originate from a different mechanism as the SGRB time distribution seems to be bimodal. This does not mean that exploding Planck stars have been detected but this raises an interesting question.

Recently, the model has been developed in \cite{hag} and the resulting phenomenology was investigated in \cite{phe1} and \cite{phe2}.

\end{document}